\documentclass[preprint]{aastex}

\usepackage{graphicx}
\usepackage{wasysym}

\shorttitle{The Origin of the Atmosphere}
\shortauthors{Dauphas}

\begin{document}

\title{The Dual Origin of the Terrestrial Atmosphere}

\author{Nicolas Dauphas}
\affil{Enrico Fermi Institute \\ The University of Chicago \\ 5640 South Ellis Avenue \\ Chicago IL 60637, USA\\ \vspace{1cm}Submitted to {\it Icarus}\\
22 September 2002\\ In press 18 June 2003\\ 31 pages, 6 tables, and 6 figures\\ \vspace{1cm}
}
\email{dauphas@uchicago.edu}

\begin{abstract}

%{\bf Abstract}

The origin of the terrestrial atmosphere is one of the most puzzling enigmas in the planetary sciences. It is suggested here that two sources contributed to its formation, fractionated nebular gases and accreted cometary volatiles. During terrestrial growth, a transient gas envelope was fractionated from nebular composition. This transient atmosphere was mixed with cometary material. The fractionation stage resulted in a high Xe/Kr ratio, with xenon being more isotopically fractionated than krypton.  Comets delivered volatiles having low Xe/Kr ratios and solar isotopic compositions. The resulting atmosphere had a near-solar Xe/Kr ratio, almost unfractionated krypton delivered by comets, and fractionated xenon inherited from the fractionation episode. The dual origin therefore provides an elegant solution to the long-standing "missing xenon" paradox. It is demonstrated that such a model could explain the isotopic and elemental abundances of Ne, Ar, Kr, and Xe in the terrestrial atmosphere.

\end{abstract}

%{\bf Keywords}

%Atmospheres, composition; Earth; Comets; Accretion; Solar Nebula

\section{Introduction}

Acquisition of a stable atmosphere is a prerequisite for the emergence and expansion of life on terrestrial planets, within the solar system and beyond. The development of life is expected to modify in turn the composition of the host atmosphere (Holland 1999), providing a planetary scale biosignature (Sagan {\it et al.} 1993). Noble gases convey important clues for deciphering the origin of the major volatile elements involved in life. Indeed, they span a large range of atomic masses, are chemically inert, and have many isotopes (Ozima and Podosek 1983).

To understand the origin of the terrestrial atmosphere one must understand which processes and which sources contributed to its formation. Despite its close proximity, the origin of the terrestrial atmosphere is one of the most puzzling enigmas in the planetary sciences (Hunten 1993, Marty and Dauphas 2002). Atmospheric noble gases (Ozima and Podosek 1983) are depleted relative to solar composition (Pepin 1991, Table 1 and Fig. 1). This depletion depends in first approximation on the mass, the light gases being more depleted and isotopically fractionated than the heavy ones. Such a trend is observed in meteorites (Mazor {\it et al.} 1970, Table 1 and Fig. 1) and it seems that the terrestrial atmosphere was derived from solar composition by some form of mass fractionation. However, this simple picture cannot explain the abundance and isotopic composition of atmospheric xenon. If the simple mass fractionation model was valid, we would expect that heavy noble gases be less depleted and less fractionated than light ones. Xenon (atomic weight 131.30) should be less depleted and less fractionated than krypton (atomic weight 83.80). Actually, the opposite is observed. Xenon is depleted in the terrestrial atmosphere by a factor of $4.8\times 10^4$ relative to solar composition while krypton is only depleted by a factor of $3.3\times 10^4$. Likewise, xenon isotopes are fractionated by 38.0 $\permil /{\rm amu}$ relative to solar while krypton isotopes are only fractionated by 7.6 $\permil /{\rm amu}$. Despite being heavier, xenon is more depleted and isotopically fractionated than krypton. This is known as the "missing xenon" paradox (Pepin 1991, Tolstikhin and O'Nions 1994). 

The Earth is not the only terrestrial planet having a "missing xenon" problem. On Mars, xenon is also more depleted and more fractionated than what is expected (Owen {\it et al.} 1977, Swindle {\it et al.} 1986, Pepin 1991). It is important to note that martian xenon was apparently derived from the solar wind component SW-Xe (Swindle and Jones 1997, Mathew {\it et al.} 1998), while terrestrial xenon was derived from a hypothetical nebular component U-Xe (Pepin and Phinney 1978, Pepin 1991, Pepin 2000). Therefore, the apparent similarity between the terrestrial and martian atmosphere should not be regarded as evidence for similarity of sources but rather as evidence for similarity of processes. 

Two categories of models have been envisioned for fractionating xenon isotopes. One assumes that the Earth accreted from porous planetesimals (Ozima and Nakazawa 1980, Zahnle {\it et al.} 1990a, Ozima and Zahnle 1993). In these planetesimals, gravitational separation of adsorbed nebular gases would have resulted in mass fractionation of trapped volatiles. This model fails to explain why light and heavy noble gases in the mantle are less fractionated than their atmospheric counterpart (Honda {\it et al.} 1991, Caffee {\it et al.} 1999). It also fails to explain why xenon in the martian atmosphere is fractionated from SW-Xe while terrestrial xenon is fractionated from U-Xe (Pepin 2000) and why unfractionated solar xenon is present in the martian mantle (Mathew and Marti 2001). Finally, it cannot reproduce the observed abundance and isotopic composition of light noble gases in the terrestrial atmosphere (Ozima and Zahnle 1993). The second category of models assumes that a massive gas envelope was lost to space on Earth or planetesimals (Zahnle and Kasting 1986, Sasaki and Nakazawa 1988, Hunten {\it et al.} 1987, Zahnle {\it et al.} 1990b, Pepin 1991, 1992, 1997, Hunten 1993). In this scenario, the rapid escape of a light gas would have exerted an aerodynamic drag on heavier gases. The balance between the upward aerodynamic drag and the downward gravitational attraction resulted in a mass-dependent loss of noble gases. A massive hydrogen atmosphere could have been acquired by capture of nebular gases or could have resulted from the reduction or photodissociation of water. The appropriate energy could have been deposited in the atmosphere as extreme ultraviolet radiation from the young evolving sun or as gravitational energy released during impacts. 

For explaining the missing xenon paradox, both models require that the fractionated gases be mixed with a source having low Xe/Kr ratio and unfractionated isotopic ratios. The two possible sources are the mantle and comets. The Xe/Kr ratio and xenon isotopic composition of the upper mantle are very close to those of the atmosphere (Moreira {\it et al.} 1998, Table 1 and Fig. 5) and it is unclear whether volatiles degassed from the Earth could have had the low Xe/Kr ratio and unfractionated isotopic ratios required by some models (Pepin 1991, 1992, 1997, Tolstikhin and O'Nions 1994). The noble gas composition of comets can be inferred from laboratory experiments (Laufer {\it et al.} 1987, Bar-Nun {\it et al.} 1988, Bar-Nun and Owen 1998, Notesco {\it et al.} 1999, Notesco {\it et al.} 2003, Owen {\it et al.} 1992, Owen and Bar-Nun 1995a, b, Table 1 and Fig. 1).  An attractive feature revealed by experiments is that noble gases trapped in amorphous water ice exhibit a clear depletion in Xe relative to Kr (Owen {\it et al.} 1992, Bar-Nun and Owen 1998). In other words, the Xe/Kr ratio is lower than the solar ratio (note that this would not be true if noble gases were trapped in crystalline ice as clathrate hydrates, Iro {\it et al.} 2003). This important observation prompted Owen and Bar-Nun (1992) and Zahnle {\it et al.} (1990a) to suggest that the atmospheres of the terrestrial planets represented the mixture of an internal component trapped in rocks and an external component delivered by icy planetesimals. Condensation of noble gases at low temperature in the protosolar nebula is not expected to create large isotopic fractionation (Pepin 1992). Laboratory experiments seem to support the view that if any, the fractionation is small for heavy noble gases (Notesco {\it et al.} 1999). It is assumed here that the noble gas isotopic composition of comets is solar.

As discussed previously, any mass fractionation on the early Earth would have resulted in a high Xe/Kr ratio and highly fractionated ratios in the residual atmosphere. An appealing possibility is that this residual atmosphere was mixed with cometary material having a low Xe/Kr ratio and unfractionated isotopic ratios (Fig. 2). The resulting atmosphere would have a near solar Xe/Kr ratio, almost unfractionated krypton, and highly fractionated xenon (Fig. 3). This is exactly what is observed on Earth and forms the "missing xenon" paradox. The idea that the terrestrial atmosphere has a dual origin (Dauphas 2003), being a mixture between a fractionated atmosphere and cometary material, is the basis of this work. Note that the results presented in this contribution critically depend on the experimentally observed depletion of xenon relative to krypton when trapped in amorphous water ice.

\section{Outline of the Model}

Ascribing a realistic environment to the fractionation episode (stage 1) is difficult. A few observations can be made that shed some light on this stage. The first observation is that, contrary to the Earth and Mars, asteroids lack the large isotopic fractionation observed for xenon (Mazor {\it et al.} 1970). This indicates that this fractionation must be closely related to the formation of terrestrial planets. Second, the mantle stable isotopic composition is distinct from, yet very close to, the atmospheric composition (Moreira {\it et al.} 1998). This suggests that active exchange occurred between the mantle and the atmosphere through degassing or recycling. Third, some radiogenic nuclides produced in the mantle are present in the atmosphere (Ozima and Podosek 1983), implying that part of the atmosphere was degassed from the mantle.  Fourth, the xenon isotopic composition of the martian atmosphere is severely fractionated while that of the martian mantle is not (Mathew and Marti 2001), supporting the view that fractionation occurred at the end of planetary accretion. Fifth, martian xenon was derived from SW-Xe while terrestrial xenon was derived from U-Xe (Swindle and Jones 1997, Mathew {\it et al.} 1998, Pepin 2000). This suggests that fractionation of xenon isotopes did not occur on planetesimals or comets. The most likely scenario is blowoff of a transient atmosphere at the end of planetary growth. Because of modeling uncertainties, it is difficult to know what is the appropriate fractionation law to apply. Instead, one may use a phenomenological model such as the generalized power law (Mar\'echal {\it et al.} 1999) which is written,
\begin{equation}
{\rm r}={\rm R}\times g^{i^n-j^n},
\end{equation}
 where r is the fractionated ratio of the nuclides $i$ and $j$, R is the initial ratio, and $g$ and $n$ are free parameters. An equivalent formulation of this law is $C^i_1/(C^i_0\,g^{i^n})=c$, with $c$ an arbitrary constant. It is assumed that, prior to mass fractionation, noble gases were present in solar proportions ($C^i_0=s\,C^i_{\odot}$, with $s$ a constant). The concentration $C_1^i$ of $i$ at the end of stage 1 can thus be related to the solar concentration  $C_{\odot}$ through,
\begin{equation}
C^i_1=x\,C^i_{\odot}\,g^{i^n},
\end{equation}
where $x$ is an arbitrary constant ($x=c\times s$). The concentration of $i$ at the end of stage 1 in the Phenomenological Model is governed by three parameters, $x$, $g$, and $n$.

After or during this fractionation episode, the Earth may have received contributions from comets (stage 2). Accretion of icy planetesimals by the Earth may have occurred in the main stage of planetary growth or might have been part of the late veneer. The late veneer refers to the material accreted by the Earth after formation of the Moon and segregation of the terrestrial core. As illustrated by the lunar cratering record, the bombardment intensity of the Earth was higher in the past than it is at present (Chyba 1990, 1991). The high and unfractionated abundances of noble metals in the terrestrial mantle (Kimura {\it et al.} 1974, Jagoutz {\it et al.} 1979) are best explained if these elements were delivered to the mantle after metal/silicate segregation. It is thus estimated that the Earth accreted  $0.7-2.7\times 10^{22}$ kg of asteroids and comets after differentiation of the core (Dauphas and Marty 2002), which is known to have occurred while the short-lived nuclide $^{182}$Hf ($t_{1/2}=9$ Ma) was still alive, within approximately 30 Ma of solar system formation (Dauphas {\it et al.} 2002a, Kleine {\it et al.} 2002, Schoenberg {\it et al.} 2002, Yin {\it et al.} 2002). The noble gas concentration of comets $C_{\circ}$ depends on the condensation temperature of ice ${\rm T}_{\circ}$. The contribution of cometary material is then,
\begin{equation}
C^i_2=y\,C^i_{\circ}({\rm T}_{\circ}),
\end{equation}
where $y$ is a scaling factor defined as ${\rm M_{\circ}/M_{\oplus}}$, M$_{\circ}$ being the mass of comets accreted by the Earth. The terrestrial atmosphere ($C^i_{a}$) is the superposition of the fractionation episode (stage 1) and the cometary accretion (stage 2),
\begin{equation}
C^i_{a}=x\,C^i_{\odot}\,g^{i^n}+y\,C^i_{\circ}({\rm T}_{\circ}).
\end{equation}
The five parameters of the model are $x$, $g$, $n$, $y$, and T$_{\circ}$. This equation can be written independently for eight nuclides. The system is overconstrained but a proper solution can be obtained by minimizing the $\chi ^2$ merit function (Press {\it et al.} 2002).

\section{Input Parameters}

Of the six noble gases He, Ne, Ar, Kr, Xe, and Rn, only four can be used to infer the origin of the terrestrial atmosphere. Indeed, He is continuously lost to space and Rn isotopes are radiogenic and radioactive with short half-lives. In addition, only two isotopes of each element provide independent information on the mass fractionation. The composition of the atmosphere can therefore be described using the following parameters for each noble gas,
\begin{equation}
F=\left[(C^i/C^j)/(C^i_{\odot}/C^j_{\odot})-1\right]/(i-j)\times 10^3,
\end{equation}
\begin{equation}
A=\log (C^i/C^i_{\odot}),
\end{equation}
where $i$ and $j$ are two isotopes of the same element and $C_{\odot}$ denotes the solar composition (see Table 1 caption for the values of the selected ratios). The values of $F$ and $A$ for Ne, Ar, Kr, and Xe and their associated uncertainties are compiled in Table~1. Graphically, forming the atmosphere can be seen as reproducing the positions ($A$) and derivatives ($F$) of the "noble gas curve" for neon, argon, krypton, and xenon in the abundance versus mass space.

At present, the dominant fraction of neon, argon, krypton, and xenon resides in the atmosphere (Dauphas and Marty 2002). Thus, the composition of the atmosphere (Ozima and Podosek 1983) approximates the composition of the Earth as a whole. Magmas formed at mid-ocean ridges bear important information on the composition of the shallow mantle.  The $^3{\rm He}$ concentration of the mantle can be calculated (Dauphas and Marty 2002, Marty and Dauphas 2003) using the observed degassing ${\rm ^{3}He}$ flux (1000 mol a$^{-1}$), the rate of oceanic crust formation (20 km$^{3}$ a$^{-1}$), and the degree of partial melting (10 \%). The mantle $^3$He concentration is thus estimated to be $1.5\times 10^{-15}$ mol g$^{-1}$. Note that this calculation depends on the assumption that the present $^3$He flux represents a reliable estimate of the time averaged flux. The abundances of the other noble gases can then be computed using elemental ratios normalized to helium (Moreira {\it et al.} 1998). Because of limited precision and large air contamination, the isotopic compositions of mantle noble gases are not yet firmly established. Within uncertainties, the isotopic compositions of argon and krypton are identical in the mantle and in the atmosphere (Marty {\it et al.} 1998, Moreira {\it et al.} 1998, Kunz 1999). The neon isotopic composition of the mantle is important as it serves as a fiducial value for deriving the composition for other noble gases (Moreira {\it et al.} 1998). Following Moreira {\it et al.} (1998) and Trieloff {\it et al.} (2000), it is assumed that the ${\rm ^{20}Ne/^{22}Ne}$ ratio of the shallow mantle is 12.5. The actual value might be higher, up to 13.8, the solar ratio (Ballentine {\it et al.} 2001, Yokochi and Marty 2003). The xenon isotopic composition can be derived from the ${\rm ^{128}Xe-^{129}Xe}$ correlation measured in CO$_2$ well gases (Caffee {\it et al.} 1999) and the inferred ${\rm ^{129}Xe}$ excess of the mantle (Moreira {\it et al.} 1998). The mantle ${\rm ^{128}Xe/^{130}Xe}$ ratio is thus estimated to be 0.476.

If solar composition was known with great accuracy, the abundances of every isotope could be used to constrain the curvature of the mass fractionation law. This idea was applied to xenon isotopes (Hunten {\it et al.} 1987) but it relies on the debatable assumption that U-Xe represents the initial nebular composition (Pepin and Phinney 1978, Pepin 1991, 2000). The enigmatic U-Xe is a nebular component distinct from the present solar composition. It has been first derived theoretically by Pepin and Phinney (1978) and it is used here as the initial atmospheric composition prior to fractionation. A recent search for its presence in achondrites was inconclusive (Busemann and Eugster 2002). 

As stated previously, it is assumed here that the isotopic composition of noble gases in comets is that of the ambient nebula. Laboratory experiments indicate that at 30 K, the normalized abundance of neon is depleted by a factor of $5\times 10^{-3}$ relative to its initial value, and that at 50 K, it is depleted by more than a factor of $10^{-5}$ (Laufer {\it et al.} 1987). The amount of argon trapped in ice follows a power dependence on the trapping temperature (Bar-Nun {\it et al.} 1988, Bar-Nun and Owen 1998, Owen and Bar-Nun 1995a). Using experimental data (Bar-Nun and Owen 1998), the solar Ar/O ratio (Anders and Grevesse 1989), and the oxygen concentration in cometary ice (Delsemme 1988, Marty and Dauphas 2002), it is straightforward to calculate the approximate argon concentration in comets, $C_{\circ}^{36}({\rm T_{\circ}})={\rm 0.055\times 10^{-T_{\circ}/11.4}}$ mol g$^{-1}$, with T$_{\circ}$ in K.  Using the calculated argon concentration and the measured enrichment factors for heavy noble gases (Owen {\it et al.} 1992, Bar-Nun and Owen 1998), it is then possible to estimate the composition of comets for Kr and Xe. The enrichment factors were measured for a finite set of trapping temperatures and they have been interpolated between measurements assuming a power dependence on T$_{\circ}$. The calculated composition of comets is given for a variety of trapping temperatures in Table 1. Recent trapping experiments indicate that the deposition rate affects the amount of gas trapped in amorphous water ice (Notesco {\it et al.} 2003), strengthening the need for measurements of noble gas concentrations in real comets.

\section{Results}

Equation (4) can be applied to two isotopes for each noble gas. There are thus eight equations in five parameters ($x$, $g$, $n$, $y$, and T$_{\circ}$). The system is overconstrained and a proper solution can only be obtained by minimizing the relative distance between the simulated and the observed atmosphere. This was done using optimization algorithms implemented in $R$ (Ihaka and Gentleman 1996). The set of parameters that minimizes the $\chi ^2$ merit function is $(x,g,n,y,{\rm T_{\circ}})=(8.158\times 10^{-15},4.701\times 10^2,0.2614,4.407\times 10^{-7},52.36)$. The isotopic and elemental compositions of the two stages involved in atmospheric formation are displayed in Fig. 2. As illustrated, neon and xenon are inherited from the fractionation episode (stage 1) while argon and krypton are delivered by cometary material (stage 2). The resulting atmosphere (Fig. 3) has near solar Xe/Kr ratio, almost unfractionated krypton, and fractionated xenon. This scenario therefore provides an elegant solution to the long standing "missing xenon" paradox. It explains the isotopic and elemental abundances of all noble gases in the terrestrial atmosphere. If U-Xe is assumed for the initial composition (identical to SW-Xe for light isotopes), then there is room in the modeled atmosphere for the subsequent addition of radiogenic $^{129}$Xe and fissiogenic $^{131-136}$Xe (Pepin 2000).

\subsection{Fractionation Episode}

For explaining the terrestrial atmosphere, the model requires that the mass fractionation be nonlinear. Interestingly, blowoff models with declining escape flux result in a concave downward curvature of the mass fractionation (Hunten {\it et al.} 1987, Pepin 1991). If the energy that drove the escape decreased with time, which would be the case if it was deposited as extreme ultraviolet radiation (Zahnle and Walker 1982, Pepin 1991) or impacts (Benz and Cameron 1990, Pepin 1997), then the escape flux must have decreased with time and the curvature follows. The timing of the fractionation episode is not precisely known but it can be constrained using xenon radiogenic isotopes. The nuclides $^{129}$Xe and $^{131-136}$Xe are the decay products of the very short-lived nuclides $^{129}$I (t$_{1/2}=16$ Ma) and $^{244}$Pu (t$_{1/2}=80$ Ma), respectively. It is thus estimated that the Earth became retentive for xenon approximately 100 Ma after solar system formation (Podosek and Ozima 2000).

The composition of the atmosphere is, in many respects, very close to the composition of the shallow mantle (Moreira {\it et al.} 1998). For instance, there seems to be a mantle "missing xenon" paradox (Section 3, Table 1). In addition, the atmosphere contains radiogenic nuclides that are only produced in the mantle (Ozima and Podosek 1983). Both observations suggest that part of the atmosphere was degassed from the mantle and possibly that the fractionation episode predated full accretion of the Earth.  It is often assumed that hydrodynamic escape occurred after the Earth had acquired its present size (Hunten {\it et al.} 1987, Sasaki and Nakazawa 1988, Pepin 1991, 1992, 1997). Loss of a massive atmosphere could also have occurred while the Earth was still accreting material, at a time when the terrestrial core was not yet differentiated and the sun was still very active. Photodissociation or reduction of the water delivered by asteroids and comets would have formed a massive hydrogen protoatmosphere. The accretional energy released by impacts would have driven the rapid escape of this transient hydrogen atmosphere. Although appealing, such a scenario is hardly amenable to modeling (Matsui and Abe 1986, Zahnle {\it et al.} 1988). In the present section, hydrodynamic escape of a transient atmosphere is simulated using the formalism developed by Hunten {\it et al.} (1987). Detailed calculations indicate that this simple formalism captures the phenomenon in its most important features (Zahnle and Kasting 1986). The reasons why blowoff of a primordial solar atmosphere is modeled is to check whether the appropriate curvature can be obtained within a realistic environment and to test how sensitive are the parameters of the accretion stage upon the fractionation episode.  

Hunten {\it et al.} (1987) developed a convenient theoretical treatment of mass fractionation during hydrodynamic escape. Advanced applications of this formalism can be found in Pepin (1991, 1992, 1997). It is assumed that the atmosphere consists of a major light constituent of mass $m$ and column density $N$, and a minor constituent $i$ of mass $m_i$ and column density $N_i$. The heavy constituent is lost to space provided that its mass is lower than a crossover mass $m_{ci}$ defined as,
\begin{equation}
m_{ci}=m+\frac{k{\rm T}\Phi}{b^igX},
\end{equation}
where $\Phi$ and $X$ are the escape flux and the mole fraction of the light constituent respectively, and $b^i$ is the diffusion parameter of $i$ (see Zahnle and Kasting 1986 and Pepin 1991 for numerical values). If the crossover mass is known for one nuclide $i$, it is then easily calculated for other nuclides,
\begin{equation}
m_{cj}=m+(m_{ci}-m)\frac{b^i}{b^j}.
\end{equation}
 It is assumed that $X$ remained constant near unity through time and that the escape flux decreased with time, $\Phi=\Phi_0\, \Psi(t)$, with $\Psi(t)$ a declining function of the time,
\begin{equation}
\Psi(t)=e^{-(t/\tau)^p},
\end{equation}
where $\tau$ is the escape timescale and $p$ is a free parameter ($p=1$ corresponds to the standard exponential decay rate, $p=2$ corresponds to a scaled normal density law, and $p \rightarrow \infty$ is the step function $\Psi =1$ for $t<\tau$ and $\Psi =0$ for $\tau <t$). The escape flux of the minor constituent can be related to the escape flux of the major constituent through,
\begin{equation}
\Phi_i=\frac{N_i}{N}\left(\frac{m_{ci}-m_i}{m_{ci}-m}\right)\Phi.
\end{equation}
Let us introduce $\mu _{i} =(m_i-m)/(m_{ci}^{0}-m)$. The rate of variation of the minor constituent $i$ can then be written as,
\begin{equation}
\frac{{\rm d}N_i}{N_i}=-\frac{\Phi_0}{N}\left[\Psi(t)-\mu _i\right]{\rm d}t,
\end{equation}
It is now assumed that the light constituent was replenished at the same rate as it was lost to space, $N(t)=N_0$. For any nuclide, loss to space began at $t=0$ and continued until the time when the crossover mass was equal to the mass of the considered nuclide. This happened at $t_{if}=\Psi^{-1}(\mu _i)$. It is then easy to integrate Eq. 11 over the proper interval to show that,
\begin{equation}
\ln \frac{N_i}{N_i^0}=\alpha \, \int_{0}^{x_{if}}\mu _i-e^{-x^p}\,{\rm d}x,
\end{equation}
where $\alpha=\Phi_0\tau /N_0$, $x$ is a nondimensional integration variable ($x=t/\tau$), and $x_{if}=[-\ln (\mu _i)]^{\frac{1}{p}}$. When $p=1$, the previous equation can be integrated analytically and takes the form  (Hunten {\it et al.} 1987, Pepin 1991),
\begin{equation}
\ln \frac{N_i}{N_i^0}=\alpha \, (\mu _i-1-\mu _i \ln \mu _i).
\end{equation}
The initial atmosphere had solar composition. The ratio $C^i_0/C^i_{\odot}$ was therefore independent of the nuclide and can be denoted $s$. The composition of the terrestrial atmosphere at the end of stage 1 is thus,
\begin{equation}
C^i_1=s\, C^i_{\odot}\, \frac{N_i}{N_i^0}.
\end{equation} 
The four parameters of the law are $s$, $\alpha$, $m^0_{cXe}$, and $p$. The second stage remains unchanged and Eq. 3 can still be applied. It was found that a large set of parameters reproduce the composition of the terrestrial atmosphere fairly well (degeneracy of the model parameters). Hunten {\it et al.} (1987) applied the very same formalism to the case of xenon alone. They limited their study to $p=1$ and concluded that $(s,\alpha, m^0_{cXe},p)=(0.633\times 10^{-3},13.13,345,1)$ could explain the fractionation and curvature of terrestrial xenon and that this set of parameters corresponded to a realistic environment for the early Earth. As illustrated in Table 3 (Hydrodynamic Model I), agreement between the observed and simulated atmosphere can be obtained for a set of parameters very close to that advocated by Hunten {\it et al.}, $(s,\alpha,m^0_{cXe},p)=(0.9561\times 10^{-3},16.04,345,1)$. The appropriate curvature of the fractionation law can therefore be obtained under realistic blowoff conditions. It is worthwhile to note that such a fractionation would also explain the curvature of the xenon isotopic composition of the terrestrial atmosphere (Hunten {\it et al.} 1987) relative to its assumed progenitor, U-Xe (Pepin and Phinney 1978). 

The success of the model is remarkable. There are however small discrepancies remaining between the simulated and the observed atmosphere. The most striking feature is the overabundance of neon in the simulated atmosphere (Hydrodynamic Model I, Table 3). Pepin (1997) was confronted with the same issue and suggested that the atmosphere prior to fractionation was depleted in neon. This is unlikely because the mantle He/Ne ratio is very close to solar (Honda {\it et al.} 1993, Moreira {\it et al.} 1998, Moreira and All\`egre 1998) and it is difficult to find a mechanism that would have depleted neon without fractionating its abundance relative to helium. As illustrated in Table 4 (Hydrodynamic Model II), a better fit to the observed atmospheric composition can be obtained by modifying the parameterization of the escape rate $(s,\alpha,m^0_{cXe},p)=(2.841\times 10^{-5},12.89,140,3.980)$. The two stages invoked in the Phenomenological (Fig. 2, Table 2) and the Hydrodynamic  Models (Tables 3 and 4) are very close.

To summarize, the fractionation episode can be successfully reproduced with a feasible set of parameters governing hydrodynamic escape of a transient atmosphere on the early Earth.

\subsection{Cometary Accretion}

How sensitive are the parameters derived for the accretion stage upon the fractionation episode? This can be investigated using the generalized power law introduced in Section 2 (Mar\'echal {\it et al.} 1999). Varying the power parameter of this law ($n$) allows to explore a wide variety of functional forms for the fractionation episode, including the exponential law (Dauphas 2003). It was found that the parameters of the accretionary stage were insensitive to the assumed form of the fractionation law. For instance, the parameters derived using the Phenomenological Model are $(y,{\rm T}_{\circ})=(4.407\times 10^{-7},52.36)$ and those derived using the blowoff model are $(y,{\rm T}_{\circ})=(4.409\times 10^{-7},53.26)$ or $(y,{\rm T}_{\circ})=(5.400\times 10^{-7},54.55)$, depending on the parameterization adopted for the escape rate, see Sections 2 and 4.1. The two parameters $y$ and T$_{\circ}$ describe the mass of comets accreted by the Earth ($5\pm 1\times 10^{-7}$ M$_{\oplus}$) and the condensation temperature of cometary ice ($50-55$ K), respectively. How do these values compare with independent estimates?

 Based on the abundances of noble metals in the terrestrial mantle, it is estimated that the Earth accreted $0.7-2.7\times 10^{22}$ kg of extraterrestrial material after segregation of the core (Dauphas and Marty 2002), which is thought to have occurred within approximately $30$ Ma of solar system formation (Dauphas {\it et al.} 2002a, Kleine {\it et al.} 2002, Schoenberg {\it et al.} 2002, Yin {\it et al.} 2002). The mass of comets accreted by the Earth is calculated to be $3.0\pm 0.6\times 10^{18}$ kg (${\rm M}_{\circ}=y\, {\rm M}_{\oplus}$). If these comets were accreted after differentiation of the core, they would only represent $10^{-4}$ by mass of the late veneer. This value agrees with the estimate, based on water deuterium to protium ratios, that comets represented less than $10^{-2}$ by mass of the impacting population (Dauphas {\it et al.} 2000). At present, comets comprise a significant fraction of Earth's impacting population, at least on the order of a few percent (Shoemaker {\it et al.} 1990). The integrated mass fraction is much lower, implying that the dynamics of asteroids and comets changed with time. Knowing the mass of comets accreted by the Earth and their composition, it is then straightforward to compute their contribution to major terrestrial volatiles. As illustrated in Table 5, the cometary contribution to the terrestrial inventory of hydrogen, carbon, and nitrogen must have been negligibly small. 

If the atmosphere was derived from mantle degassing, this implies that comets were not delivered as a late veneer but were accreted during the main stage of planetary accretion. Note that cometary accretion could have been contemporaneous with fractionation of a transient atmosphere (Section 4.1). 

Following different lines of evidence, Owen {\it et al.} (1992) estimated that the formation temperature of the comets that impacted the Earth must have been around 50 K, very close to the value previously derived. This is not unexpected because the "missing xenon" paradox requires a source with a low Xe/Kr ratio, and the cometary Xe/Kr ratio is subsolar for a narrow range of trapping temperatures centered around 50 K (Owen {\it et al.} 1992, Bar-Nun and Owen 1998). Simulation of cometary trajectories in the early solar system indicates that the comets that impacted the Earth must have originated beyond the orbit of Uranus (Morbidelli {\it et al.} 2000), in a region of the nebula where the temperature was probably lower than 70 K (Yamamoto 1985). This upper-limit is consistent with the value advocated in the present contribution. 

As discussed in Section 3, trapping conditions such as the deposition rate affect the amount of gas trapped in amorphous water ice (Notesco {\it et al.} 2003). The mass and temperature of comets derived here using high deposition rates (Owen {\it et al.} 1992, Bar-Nun and Owen 1998) probably represent upper-limits.

\section{Origin of Water on Earth}

The contribution of comets formed below 55 K to the terrestrial inventory of major volatiles must have been negligibly small (Table 5). This result agrees with the marked difference between the deuterium to protium ratio of the Earth and that of comets (Balsiger {\it et al.} 1995, Eberhardt {\it et al.} 1995, Meier {\it et al.} 1998, Bockel\'ee-Morvan {\it et al.} 1998, Dauphas {\it et al.} 2000), and the noble gas and noble metal concentrations of the Earth (Dauphas and Marty 2002). If comets did not deliver major volatiles, how did the Earth acquire its oceans? 

Water could have been delivered by a late asteroidal veneer (Dauphas {\it et al.} 2000). Some meteorite groups contain large amounts of water with the appropriate D/H ratio (Kerridge 1985). Primitive carbonaceous chondrites contain 5-10 \% water. The mass of the late veneer accreted after core formation is $0.7-2.7\times 10^{22}$ kg. Hence, accretion of asteroidal material could have delivered $0.3-2.7\times 10^{21}$ kg of water. This range encompasses the mass of terrestrial oceans, $1.4\times 10^{21}$ kg. Although appealing, this idea has difficulties. First, the osmium isotopic composition of the mantle is different from that of carbonaceous chondrites (Meisel {\it et al.} 1996, 2001, Walker {\it et al.} 2001). If noble metals in the terrestrial mantle were entirely delivered by the late veneer, this would imply that carbonaceous asteroids did not impact the Earth in a late accretionary stage. Note however that the osmium isotopic composition of the mantle is reconcilable with a late carbonaceous veneer, if noble metals delivered to the mantle were mixed with the fractionated residue left over after core formation (Dauphas {\it et al.} 2002b). The recent finding in meteorites of ruthenium isotope anomalies (Chen {\it et al.} 2003), an element which was essentially all delivered to the Earth after core formation, may ultimately help to solve the question of the nature of the late veneer. A second difficulty arises with noble gases. Indeed, accretion by the Earth of water from a late carbonaceous veneer would have probably delivered too much xenon of inappropriate isotopic composition (Owen and Bar-Nun 1995a, Dauphas {\it et al.} 2000). 

Owen and Bar-Nun (1995a, b) and Delsemme (1999) suggested that comets formed in the Jupiter region of the nebula delivered major volatiles. These comets would have formed at comparatively high temperature (100 K). As a consequence, they would have trapped less noble gases and would exhibit a lower D/H ratio due to exchange with protosolar hydrogen. Xenon would certainly be depleted in such comets but it seems unlikely that this depletion attained the required level for delivering water without disturbing noble gases. Morbidelli {\it et al.} (2000) modeled comet trajectories in the nascent solar system. They concluded that the contribution of cometesimals from the Jupiter zone to the terrestrial water inventory was negligible.

Unless one calls on hypothetical impactors having high H$_2$O/Xe ratios, late accretion by the Earth of asteroidal and cometary material does not provide a straightforward explanation to the origin of the terrestrial oceans. This may indicate that major volatiles and noble gases were decoupled at some stage of planetary formation. Hydrogen, carbon, and nitrogen form reactive compounds that can interact with the geosphere. During or at the end of planetary growth, major volatiles could have been trapped in the Earth (Matsui and Abe 1986, Morbidelli {\it et al.} 2000, Abe {\it et al.} 2000) while inert noble gases would have accumulated in a transient atmosphere. This transient atmosphere would have been lost to space as a result of impact erosion and continuous escape. Differential retention of gases during planetary growth therefore provides an elegant solution to the observed decoupling between inert and reactive gases, and explains why the terrestrial H$_2$O/Xe ratio is so high.   

\section{Mantle Noble Gases}

When and where did the fractionation and accretion stages occur? Although debated, the composition of the mantle bears important information on this issue. The following discussion relies on an incomplete and uncertain description of the mantle composition and noble gas behavior. It is therefore highly speculative and should be regarded as such. The silicate Earth can be divided into two reservoirs (All\`egre {\it et al.} 1986/87, Marty and Dauphas 2002). One was severely outgassed, resulting in highly radiogenic isotopic ratios. This reservoir is sampled at mid-ocean ridges and represents the shallow part of the mantle. The other was much less outgassed and is sampled by ascending mantle plumes from the deep Earth. The composition of the shallow mantle is well documented while that of the deep Earth is to a large extent unknown. The shallow mantle apparently exhibits the same "missing xenon" paradox as the atmosphere (Section 3, Table 1). This is best explained if the atmosphere was derived in part from mantle degassing or if atmospheric volatiles were recycled in the mantle. 

The view that part of the atmosphere was degassed from the mantle is supported by the presence in the atmosphere of radiogenic nuclides that are produced in the silicate Earth, $^{4}$He, $^{21}$Ne, $^{40}$Ar, $^{129}$Xe, and $^{131-136}$Xe. Conversely, there is growing evidence that heavy noble gases are efficiently recycled in the mantle. Sarda {\it et al.} (1999) observed a correlation between the argon and lead isotopic compositions in mid-Atlantic ridge basalts. The most straightforward interpretation is that Ar is recycled at subduction zones (see Burnard 1999 for a different explanation). An additional piece of evidence is given by Matsumoto {\it et al.} (2001) who observed a correlation in xenoliths between the abundances of mantle derived $^3$He and atmospheric $^{36}$Ar. If argon is indeed recycled, then this must be true for heavier noble gases, krypton and xenon.   

Although very similar, the composition of the shallow mantle is not identical to the composition of the atmosphere. The most striking difference is the neon isotopic composition of the mantle (Honda {\it et al.} 1991, Trieloff {\it et al.} 2000), which is very close to solar composition. This observation, together with the fact that the Ne/He ratio of the shallow mantle is close to that of the sun (Honda {\it et al.} 1993, Moreira {\it et al.} 1998, Moreira and All\`egre 1998), requires the presence of an unfractionated solar component in the Earth. The second important difference is the xenon stable isotopic composition, which is apparently less fractionated in the mantle than it is in the atmosphere (Caffee {\it et al.} 1999).

The model advocated for the formation of the terrestrial atmosphere postulates that it is derived from solar composition. If atmospheric volatiles were recycled in the mantle and mixed with solar gases trapped in the silicate Earth, this would naturally explain the fact that light noble gases share affinities with the sun while heavy noble gases share affinities with the atmosphere (Fig. 4). A complication arises with xenon, which appears to have distinct isotopic compositions in the mantle and in the atmosphere. If confirmed, this observation would require the presence of a meteoritic component in the mantle. Note that fractionation of noble gases during recycling cannot explain the xenon isotopic composition of the mantle because recycled Xe is expected to be enriched in the heavy isotopes, not depleted. If atmospheric noble gases were recycled in the mantle, the fractionation episode introduced in Section 2 could have occurred on a fully accreted planet and cometary material could have been delivered as a late veneer. It is assumed here that recycled atmospheric volatiles were mixed with solar and possibly meteoritic gases trapped in the mantle,
\begin{equation}
C^i_m=x\,C^{i*}_a+y\,C^i_{\odot}+z\,C^i_{\bullet},
\end{equation}
where $C^{i*}_a$ represents the atmospheric composition fractionated in subduction zones (for simplicity, it is assumed here that there is no fractionation), $C^i_{\bullet}$ represents the meteoritic component (required for explaining the xenon isotopic composition of the mantle), and $x$, $y$, and $z$ are scaling factors. The set of parameters that minimizes the $\chi ^2$ merit function is $(x,y,z)=(7.730\times 10^{-11},1.853\times 10^{-6},5.158\times 10^{-3})$ (Table 6). As illustrated in Fig. 5, a good agreement is obtained between the observed and the modeled mantle composition. A better fit would be obtained if noble gases were fractionated during recycling, the heavy ones being preferentially recycled relative to the light ones (Fig. 4 and 5). Such a fractionation is very likely to occur. As far as radiogenic isotopes are concerned, a formalism similar to that developed by Porcelli and Wasserburg (1995a, b) would probably apply.

Note that if recycling of noble gases does not occur, then the atmosphere could have been derived from degassing of the upper part of a stratified mantle, and the distinct solar signature of the present silicate Earth could have been acquired by subsequent homogenization of the initial layered mantle. If so, then Eq. 15 would still apply.

\section{Perspectives}

The origin of the terrestrial atmosphere is one of the most puzzling enigmas of the planetary sciences. One of the most difficult features to understand is the "missing xenon" paradox. Being heavier, xenon should be less depleted and less fractionated than krypton. Actually, the opposite is observed. Any fractionation event on the early Earth would have resulted in high Xe/Kr and highly fractionated isotopic ratios. Interestingly, noble gases trapped in cometary ice around 50 K exhibit a clear depletion in Xe relative to Kr (Owen {\it et al.} 1992, Bar-Nun and Owen 1998). An appealing possibilty is that the terrestrial atmosphere has a dual origin, being a mixture between a fractionated atmosphere and cometary material. The resulting atmosphere would have near solar Xe/Kr, almost unfractionated krypton delivered by comets, and fractionated xenon inherited from the fractionation episode. It is demonstrated that such a model (Fig. 6) can account for the observed elemental and isotopic compositions of all noble gases in the terrestrial atmosphere. The appropriate concave downward curvature of the fractionation law can be obtained under realistic blowoff conditions for the early Earth (Hunten {\it et al.} 1987). The trapping temperature of noble gases is estimated to be lower than $55$ K, which is consistent with simulation indicating that the comets accreted by the Earth formed beyond the orbit of Uranus (Morbidelli {\it et al.} 2000). The mass of comets accreted by the Earth ($3\times 10^{18}$ kg) represents a tiny fraction ($<0.0001$) of the late veneer accreted after differentiation of the core (Dauphas and Marty 2002).

The model relies on experiments for deriving the composition of noble gases trapped in cometary ice. The processes, timescales, and location of cometary formation are still unknown.  Measurement of the noble gas composition of real comets will therefore provide key information on the origin of the terrestrial atmosphere.

\acknowledgments
I wish to thank R.O. Pepin and F. Robert for their thoughtful reviews of this paper. I also wish to acknowledge R. Yokochi,  A.M. Davis, B. Marty, and T. Owen for fruitful discussions. This work was supported by the National Aeronautics and Space Administration, through grant NAGW-9510 (to A.M. Davis).

\newpage

{\bf References}

Abe, Y., E. Ohtani, T. Okuchi, K. Righter, and M. Drake 2000. Water in the early Earth. In {\it Origin of the Earth and Moon} (R.M Canup, and K. Righter, Eds.), pp. 413-433. The University of Arizona Press, Tucson.

All\`egre, C.J., T. Staudacher, and P. Sarda 1986/87. Rare gas systematics: formation of the atmosphere, evolution and structure of the earth's mantle. {\it Earth Planet. Sci. Lett.} {\bf 81}, 127-150.

Anders, E., and N. Grevesse 1989. Abundances of the elements: meteoritic and solar. {\it Geochim. Cosmochim. Acta} {\bf 53}, 197-214.

%Anders, E., and E. Zinner 1993. Interstellar grains in primitive meteorites: diamond, silicon carbide, and graphite. {\it Meteoritics} {\bf 28}, 490-514.

Ballentine, C.J., D. Porcelli, and R. Wieler 2001. Noble gases in mantle plumes. {\it Science} {\bf 291}, 2269a

Balsiger, H., K. Altwegg, and J. Geiss 1995. D/H and ${\rm ^{18}O/^{16}O}$ ratio in the hydronium ion and in neutral water from in situ ion measurements in comet Halley. {\it J. Geophys. Res.} {\bf 100}, 5827-5834.

Bar-Nun, A., I. Kleinfeld, and E. Kochavi 1988. Trapping of gas mixtures by amorphous water ice. {\it Phys. Rev. B.}, {\bf 38}, 7749-7754.

Bar-Nun, A., and T. Owen 1998. Trapping of gases in water ice and consequences to comets and the atmospheres of the inner planets. In {\it Solar System Ices} (B.B. Schmitt, C. De Bergh, and M. Festou, Eds.), pp. 353-366. Kluwer Academic Publishers, Dordrecht.

Benz, W., and A.G.W. Cameron 1990. Terrestrial effects of the giant impact. In {\it Origin of the Earth} (H.E. Newsom and J.H. Jones, Eds.), pp. 61-67. Oxford Univ. Press, New York.

Bockel\'ee-Morvan, D., D. Gautier, D.C. Lis, K. Young, J. Keene, T. Phillips, T. Owen, J. Crovisier, P.F. Goldsmith, E.A. Bergin, D. Despois, and A. Wootten 1998. Deuterated water in comet C/1996 B2 (Hyakutake) and its implications for the origin of comets. {\it Icarus} {\bf 133}, 147-162.

Burnard, P.G. 1999. Origin of argon-lead isotopic correlation in basalts. {\it Science} {\bf 286}, 871a.

Busemann, H., and O. Eugster 2002. The trapped noble gas component in achondrites. {\it Meteoritics Planet. Sci.} {\bf 37}, 1865-1891.

Caffee, M.W., G.B. Hudson, C. Velsko, G.R. Huss, E.C. Alexander Jr., and A.R. Chivas 1999. Primordial noble gases from Earth's mantle: identification of a primitive component. {\it Science} {\bf 285}, 2115-2118.

Chen, J.H., D.A. Papanastassiou, and G.J. Wasserburg 2003. Endemic Ru isotopic anomalies in iron meteorites and in Allende. {\it Lunar Planet. Sci.} {\bf XXXIV}, \#1789.

Chyba, C.F. 1990. Impact delivery and erosion of planetary oceans in the early inner solar system. {\it Nature} {\bf 343}, 129-133.

Chyba, C.F. 1991. Terrestrial mantle siderophiles and the lunar impact record. {\it Icarus} {\bf 92}, 217-233.

Dauphas, N. 2003. The origin of the terrestrial atmosphere: early fractionation and cometary accretion. {\it Lunar Planet. Sci.} {\bf XXXIV}, \#1813. 

Dauphas N., and B. Marty 2002. Inference on the nature and the mass of Earth's late veneer from noble metals and gases. {\it J. Geophys. Res.}, {\bf 107} (E12), 5129, doi:10.1029/2001JE001617.

Dauphas, N., B. Marty, and L. Reisberg 2002a. Inference on terrestrial genesis from molybdenum isotope systematics. {\it Geophys. Res. Lett.} {\bf 29} (6), 1084, doi:10.1029/2001GL014237.

%Dauphas, N., L. Reisberg, and B. Marty 2001. Solvent extraction, ion chromatography, and mass spectrometry of molybdenum isotopes. {\it Anal. Chem.} {\bf 73}, 2613-2616.

Dauphas, N., L. Reisberg, and B. Marty 2002b. An alternative explanation for the distribution of highly siderophile elements in the Earth. {\it Geochem. J.} {\bf 36}, 409-419.

Dauphas, N., F. Robert, and B. Marty 2000. The late asteroidal and cometary bombardment of Earth as recorded in water deuterium to protium ratio. {\it Icarus} {\bf 148}, 508-512.

Delsemme, A.H. 1988. The chemistry of comets. {\it Phil. Trans. R. Soc. Lond.} {\bf A325}, 509-523.

Delsemme, A.H. 1999. The deuterium enrichment observed in recent comets is consistent with the cometary origin of seawater. {\it Planet. Space Sci.} {\bf 47}, 125-131.

Eberhardt, P., M. Reber, D. Krankowsky, and R.R. Hodges 1995. The ${\rm D/H}$ and ${\rm ^{18}O/^{16}O}$ ratios in water from comet P/Halley. {\it Astron. Astrophys.} {\bf 302}, 301-316.

%Fernandez, J.A., and W.H. Ip 1981. Dynamical evolution of a cometary swarm in the outer planetary region. {\it Icarus} {\bf 47}, 470-479.

Holland, H.D. 1999. When did the Earth's atmosphere become oxic? A reply. {\it Geochemical News} {\bf 100}, 20-22.

Honda, M., I. McDougall, D.B. Patterson, A. Doulgeris, and D.A. Clague 1991. Possible solar noble-gas component in Hawaiian basalts. {\it Nature} {\bf 349}, 149-151.

Honda, M., I. McDougall, D.B. Patterson, A. Doulgeris, and D.A. Clague 1993. Noble gases in submarine pillow basalt glasses from Loihi and Kilauea, Hawaii: A solar component in the Earth. {\it Geochim. Cosmochim. Acta} {\bf 57}, 859-874.

Hunten, D.M. 1993. Atmospheric evolution of the terrestrial planets. {\it Science} {\bf 259}, 915-920.

Hunten, D.M., R.O. Pepin, and J.C.G. Walker 1987. Mass fractionation in hydrodynamic escape. {\it Icarus} {\bf 69}, 532-549.

Ihaka, R., and R. Gentleman 1996. R: a language for data analysis and graphics. {\it J. Comput. Graph. Stat.} {\bf 5}, 299-314.

Iro, N., D. Gautier, F. Hersant, D. Bockel\'ee-Morvan. and J.I. Lunine 2003. An interpretation of the nitrogen deficiency in comets. {\it Icarus} {\bf 161}, 511-532.

Jagoutz, E., H. Palme, H. Baddenhausen, K. Blum, M. Cendales, G. Dreibus, B. Spettel, V. Lorenz, and H. W\"anke 1979. The abundances of major, minor and trace elements in the Earth's mantle as derived from primitive ultramafic nodules. {\it Lunar Planet. Sci.} {\bf X}, 2031-2050.

Kerridge, J.F. 1985. Carbon, hydrogen and nitrogen in carbonaceous chondrites: abundances and isotopic compositions in bulk samples. {\it Geochim. Cosmochim. Acta} {\bf 49}, 1707-1714.

Kimura, K., R.S. Lewis, and E. Anders 1974. Distribution of gold and rhenium between nickel-iron and silicate melts: implications for the abundance of siderophile elements on the Earth and Moon. {\it Geochim. Cosmochim. Acta} {\bf 38}, 683-701.

Kleine, T., C. M\"unker, K. Mezger, and H. Palme 2002. Rapid accretion and early core formation on asteroids and the terrestrial planets from Hf-W chronometry. {\it Nature} {\bf 418}, 952-955.

Kunz, J. 1999. Is there solar argon in the Earth's mantle? {\it Nature} {\bf 399}, 649-650.

%Kunz, J., T. Staudacher, and C.J. All\`egre 1998. Plutonium-fission xenon found in Earth's mantle. {\it Science} {\bf 280}, 877-880.

Laufer, D., E. Kochavi, and A. Bar-Nun 1987. Structure and dynamics of amorphous water ice. {\it Phys. Rev. B.} {\bf 36}, 9219-9228.

Mar\'echal, C.N., P. T\'elouk, and F. Albar\`ede 1999. Precise analysis of copper and zinc isotopic compositions by plasma-source mass spectrometry. {\it Chem. Geol.} {\bf 156}, 251-273.

Marty, B., I. Tolstikhin, I.L. Kamensky, V. Nivin, E. Balaganskaya, and J.-L. Zimmermann 1998. Plume-derived rare gases in 380 Ma carbonatites from the Kola region (Russia) and the argon isotopic composition in the deep mantle. {\it Earth Planet. Sci. Lett.} {\bf 164}, 179-192. 

Marty, B., and N. Dauphas 2002. Formation and early evolution of the atmosphere. In {\it Early Earth: Physical, Chemical and Biological Development} (C.M.R. Fowler, C.J. Ebinger, and C.J. Hawkesworth, Eds.), {\it Geological Society Special Publications} {\bf 199}, 213-229.

Marty, B., and N. Dauphas 2003. The nitrogen record of crust-mantle interaction from Archean to Present. {\it Earth Planet. Sci. Lett.} {\bf 206}, 397-410.

Mathew, K.J., J.S. Kim, and K. Marti 1998. Martian atmospheric and indigenous components of xenon and nitrogen in the Shergotty, Nakhla, and Chassigny group meteorites. {\it Meteoritics Planet. Sci.} {\bf 33}, 655-664.

Mathew, K.J., and K. Marti 2001. Evolution of martian volatiles: nitrogen and noble gas components in ALH84001 and Chassigny. {\it J. Geophys. Res.} {\bf 106}, 1401-1422.

Matsui, T., and Y. Abe 1986. Evolution of an impact-induced atmosphere and magma ocean on the accreting Earth. {\it Nature} {\bf 319}, 303-305.

Matsumoto, T., Y. Chen, and J.I. Matsuda 2001. Concomitant occurrence of primordial and recycled noble gases in the Earth's mantle. {\it Earth Planet. Sci. Lett.} {\bf 185}, 35-47.

Mazor, E., D. Heymann, and E. Anders 1970. Noble gases in carbonaceous chondrites. {\it Geochim. Cosmochim. Acta} {\bf 34}, 781-824.

Meier, R., T.C. Owen, H.E. Matthews, D.C. Jewitt, D. Bockel\'ee-Morvan, N. Biver, J. Crovisier, and D. Gautier 1998. A determination of the HDO/H$_2$O ratio in comet C/1995 O1 (Hale-Bopp). {\it Science} {\bf 279}, 842-844.

Meisel, T., R.J. Walker, and J.W. Morgan 1996. The osmium isotopic composition of the earth's primitive upper-mantle. {\it Nature} {\bf 383}, 517-520.

Meisel, T., R.J. Walker, A.J. Irving, and J.-P. Lorand 2001. Osmium isotopic compositions of mantle xenoliths: a global perspective. {\it Geochim. Cosmochim. Acta} {\bf 65}, 1311-1323.

Morbidelli, A., J. Chambers, J.I. Lunine, J.M. Petit, F. Robert, G.B. Valsecchi, and K.E. Cyr 2000. Source regions and timescales for the delivery of water to Earth. {\it Meteoritics Planet. Sci.} {\bf 35}, 1309-1320.

Moreira, M., and C.J. All\`egre 1998. Helium-neon systematics and the structure of the mantle. {\it Chem. Geol.} {\bf 147}, 53-59.

Moreira, M., J. Kunz, and C. All\`egre 1998. Rare gas systematics in popping rock: isotopic and elemental compositions in the upper mantle. {\it Science} {\bf 279}, 1178-1181.

Notesco, G., D. Laufer, A. Bar-Nun, and T. Owen 1999. An experimental study of the isotopic enrichment in Ar, Kr, and Xe when trapped in water ice. {\it Icarus} {\bf 142}, 298-300.

Notesco, G., A. Bar-Nun, and T. Owen 2003. Gas trapping in water ice at very low deposition rates and implications for comets. {\it Icarus} {\bf 162}, 183-189.

Owen, T., and A. Bar-Nun 1995a. Comets, impacts, and atmospheres. {\it Icarus} {\bf 116}, 215-226.

Owen, T., and A. Bar-Nun 1995b. Comets, impacts, and atmospheres II. Isotopes and noble gases. In {\it Volatiles in the Earth and Solar System} (K. Farley, Ed.), pp. 123-138. American Institute of Physics, New York.

Owen, T., A. Bar-Nun, and I. Kleinfeld 1992. Possible cometary origin of heavy noble gases in the atmospheres of Venus, Earth and Mars. {\it Nature} {\bf 358} 43-46.

Owen, T., K. Biemann, D.R. Rushneck, J.E. Biller, D.W. Howarth, and L. Lafleur 1977. The composition of the atmosphere at the surface of Mars. {\it J. Geophys. Res.} {\bf 82}, 4635-4639.

Ozima, M., and K. Nakazawa 1980. Origin of rare gases in the Earth. {\it Nature} 284, 313-316.

Ozima, M., and F.A. Podosek 1983. {\it Noble gas geochemistry}. Cambridge Univ. Press, Cambridge.

Ozima, M., and K. Zahnle 1993. Mantle degassing and atmospheric evolution: noble gas view. {\it Geochem. J.} {\bf 27}, 185-200.

Pepin, R.O. 1991. On the origin and early evolution of terrestrial planet atmospheres and meteoritic volatiles. {\it Icarus} {\bf 92}, 2-79.

Pepin, R.O. 1992. Origin of noble gases in the terrestrial planets. {\it Annu. Rev. Earth Planet. Sci.} {\bf 20}, 389-430.

Pepin, R.O. 1997. Evolution of earth's noble gases: consequences of assuming hydrodynamic loss driven by giant impact. {\it Icarus} {\bf 126}, 148-156.

Pepin, R.O. 2000. On the isotopic composition of primordial xenon in terrestrial planet atmospheres. {\it Space Sci. Rev.} {\bf 92}, 371-395.

Pepin, R.O., and D. Phinney 1978. Components of xenon in the solar system. {\it Unpublished preprint}.

Pepin, R.O., R.H. Becker, and P.E. Rider 1995. Xenon and krypton isotopes in extraterrestrial regolith soils and in the solar wind. {\it Geochim. Cosmochim. Acta} {\bf 59}, 4997-5022.

Pepin, R.O., R.H. Becker, and D.J. Schlutter 1999. Irradiation records in regolith materials. I: Isotopic compositions of solar-wind neon and argon in single lunar mineral grains. {\it Geochim. Cosmochim. Acta} {\bf 63}, 2145-2162.

Podosek, F.A., and M. Ozima 2000. The xenon age of the Earth. In {\it Origin of the Earth and Moon} (R.M Canup, and K. Righter, Eds.), pp. 63-72. The University of Arizona Press, Tucson.

Porcelli, D., and G.J. Wasserburg 1995a. Mass transfer of xenon through a steady-state upper mantle. {\it Geochim. Cosmochim. Acta} {\bf 59} 1991-2007.

Porcelli, D., and G.J. Wasserburg 1995b. Mass transfer of helium, neon, argon, and xenon through a steady-state upper mantle. {\it Geochim. Cosmochim. Acta} {\bf 59} 4921-4937.

Press, W.H., S.A. Teukolsky, W.T. Vetterling, and B.P. Flannery 2002. {\it Numerical Recipes in C}. Cambridge Univ. Press, Cambridge.

Sagan, C., W.R. Thompson, R. Carlson, D. Gurnett, and C. Hord 1993. A search for life on Earth from the Galileo spacecraft. {\it Nature} {\bf 365} 715-721.

Sarda, P., M. Moreira, and T. Staudacher 1999. Argon-lead isotopic correlation in mid-atlantic ridge basalts. {\it Science} {\bf 283}, 666-668.

Sasaki, S., and K. Nakazawa 1988. Origin of isotopic fractionation of terrestrial Xe: hydrodynamic fractionation during escape of the primordial H$_2$--He atmosphere. {\it Earth Planet. Sci. Lett.} {\bf 89}, 323-334.

Schoenberg, R., B.S. Kamber, K.D. Collerson, and O. Eugster 2002. New W-isotope evidence for rapid terrestrial accretion and very early core formation. {\it Geochim. Cosmochim. Acta} {\bf 66}, 3151-3160.

Shoemaker, E.M., R.F. Wolfe, and C.S. Shoemaker 1990. Asteroid and comet flux in the neighborhood of Earth. In {\it Global Catastrophes in Earth History} (V.L. Sharpton, and P.O. Ward, Eds.), pp. 155-170. Geological Society of America, Boulder. 

Swindle, T.D., M.W. Caffee, and C.M. Hohenberg 1986. Xenon and other noble gases in shergottites. {\it Geochim. Cosmochim. Acta} {\bf 50}, 1001-1015.

Swindle, T.D., and J.H. Jones 1997. The xenon isotopic composition of the primordial martian atmosphere: contributions from solar and fission components. {\it J. Geophys. Res.} {\bf 102}, 1671-1678.

Tolstikhin, I.N., and R.K. O'Nions 1994. The earth's missing xenon: a combination of early degassing and of rare gas loss from the atmosphere. {\it Chem. Geol.} {\bf 115}, 1-6.

Trieloff, M., J. Kunz, D.A. Clague, D. Harrison, and C.J. All\`egre 2000. The nature of pristine noble gases in mantle plumes. {\it Science} {\bf 288}, 1036-1038.

Walker, R.J., M.F. Horan, J.W. Morgan, and T. Meisel 2001. Osmium isotopic compositions of chondrites and Earth's primitive upper mantle: constraints on the late veneer. {\it Lunar Planet. Sci.} {\bf XXXII}, \# 1152.

Wieler, R. 1998. The solar noble gas record in lunar samples and meteorites. {\it Space Sci. Rev.} {\bf 85}, 303-314.

Wieler, R., and H. Baur 1994. Krypton and xenon from the solar wind and solar energetic particles in two ilmenites of different antiquity. {\it Meteoritics} {\bf 29}, 570-580.

Yamamoto, T. 1985. Formation environment of cometary nuclei in the primordial solar nebula. {\it Astron. Astrophys.} {\bf 142}, 31-36.

Yokochi, R., and B. Marty 2003. A determination of the neon isotopic composition of the mantle. {\it Geophysical Research Abstracts} {\bf 5}, 06823.

Yin, Q., S.B. Jacobsen, K. Yamashita, J. Blichert-Toft, P. T\'elouk, and F. Albar\`ede 2002. A short timescale for terrestrial planet formation from Hf-W chronometry of meteorites. {\it Nature} {\bf 418}, 949-952.

Zahnle, K.J., and J.C.G. Walker 1982. The evolution of solar ultraviolet luminosity. {\it Rev. Geophys. Space Phys.} {\bf 20}, 280-292.

Zahnle, K.J., and J.F. Kasting 1986. Mass fractionation during transonic escape and implications for loss of water from Mars and Venus. {\it Icarus} {\bf 68}, 462-480.

Zahnle, K., J.F. Kasting, and J.B. Pollack 1988. Evolution of a steam atmosphere during Earth's accretion. {\it Icarus} {\bf 74}, 62-97.

Zahnle, K., J.B. Pollack, and J.F. Kasting 1990a. Xenon fractionation in porous planetesimals. {\it Geochim. Cosmochim. Acta.} {\bf 54}, 2577-2586.

Zahnle, K., J.F. Kasting, and J.B. Pollack 1990b. Mass fractionation of noble gases in diffusion-limited hydrodynamic hydrogen escape. {\it Icarus} {\bf 84}, 502-527.

\newpage

\clearpage
\newpage

\begin{deluxetable}{lrcccccc}
\tabletypesize{\tiny}
\tablewidth{0pt}
\tablecaption{Input Parameters}
\tablehead{
\colhead{} & \colhead{} & \colhead{} & \colhead{} & \colhead{${\rm Ne}$} & \colhead{${\rm Ar}$} & \colhead{${\rm Kr}$} & \colhead{${\rm Xe}$}}
\startdata

Solar & $F$ & & & 0 & 0 & 0 & 0 \\
& $A$      & & & 0 & 0 & 0 & 0 \\
\\
Atmosphere & $F$ & & & $144.9$ & $25.0$ & $7.6$ & $36.2$ \\ 
	   & $\pm$    & & & $3.6$ & $8.9$ & $1.3$ & $1.6$ \\
& $A$      & & & $-8.21$ & $-6.35$ & $-4.52$ & $-4.68$ \\
	   & $\pm$    & & & $0.10$ & $0.10$ & $0.07$ & $0.08$ \\	   
\\
Mantle & $F$ & & & 47.1 & 25.0 & 7.6 & 31.8 \\
	   & $\pm$    & & & $8.8$ & $8.9$ & $1.3$ & $2.4$ \\
       & $A$      & & & -10.33 & -8.76 & -6.77 & -6.41 \\ 
	   & $\pm$    & & & $0.20$ & $0.20$ & $0.20$ & $0.20$ \\
\\
Comets (T$_{\circ}$) & $F$ & & & 0 & 0 & 0 & 0 \\
& $A$  & {\it 30}  & & -0.7 & 1.78 & 1.78 & 1.78 \\
&      & {\it 35}  & & $<-0.7$ & 1.33 & 2.09 & 1.87 \\
&	& {\it 40}  & & $<-0.7$ & 0.90 & 2.19 & 1.81 \\
&	& {\it 45}  & & $<-0.7$ & 0.46 & 1.98 & 1.56 \\
&	& {\it 50}  & & $<-3.4$ & 0.02 & 1.77 & 1.30 \\
&	& {\it 55}  & & $<-3.4$ & -0.42 & 1.65 & 1.07 \\
&	& {\it 60}  & & $<-3.4$ & -0.86 & 1.71 & 1.06 \\
&	& {\it 65}  & & $<-3.4$ & -1.29 & 1.44 & 1.11 \\
&	& {\it 70}  & & $<-3.4$ & -1.73 & 1.17 & 1.16 \\
\\
Asteroids & $F$ & & & 177.5 & 26.8 & 7.4 & 1.0 \\
& $A$ & & & -6.75 & -4.79 & -3.19 & -1.98 \\

\enddata
\tablecomments{See Section 3 for formal definitions of $F$ and $A$. Abundances are normalized to the solar composition, ${\rm ^{20}Ne}=8.107\times 10^{-5}$, ${\rm ^{36}Ar}=2.153\times 10^{-6}$, ${\rm ^{84}Kr}=6.511\times 10^{-10}$, and ${\rm ^{130}Xe}=5.194\times 10^{-12}$ mol g$^{-1}$ (Anders and Grevesse 1989). Isotopic ratios are also normalized to the solar composition, ${\rm ^{20}Ne/^{22}Ne=13.8\pm 0.1}$, ${\rm ^{36}Ar/^{38}Ar=5.6\pm 0.1}$ (Wieler 1998, Pepin {\it et al.} 1999), ${\rm ^{83}Kr/^{84}Kr=0.20291\pm 0.00026}$, and ${\rm ^{128}Xe/^{130}Xe=0.5083\pm 0.0006}$ (Pepin 2000, Pepin {\it et al.} 1995, Wieler and Baur 1994). The trapping temperature of comets is in K. The atmospheric concentration is reported per gram of the Earth.  The mantle concentration is that of the shallow mantle feeding mid-ocean ridges. Uncertainties affecting the atmospheric and mantle compositions are due to a large extent to normalization to the solar composition. Uncertainties are $2\sigma$. References: Solar- Anders and Grevesse (1989), Pepin (2000), Pepin {\it et al.} (1995, 1999), Wieler (1998), Wieler and Baur (1994). Atmosphere- Ozima and Podosek (1983). Mantle- Moreira {\it et al.} (1998), Trieloff {\it et al.} (2000), Caffee {\it et al.} (1999), Kunz (1999). Comets- Anders and Grevesse (1989), Bar-Nun {\it et al.} (1988), Bar-Nun and Owen (1998), Delsemme (1988), Marty and Dauphas (2002). Asteroids- Mazor {\it et al.} (1970).
}
\end{deluxetable}

\clearpage

\begin{deluxetable}{lrccccc}
\tabletypesize{\scriptsize}
\tablewidth{0pt}
\tablecaption{Phenomenological Model}
\tablehead{
\colhead{} & \colhead{} & \colhead{} & \colhead{${\rm Ne}$} & \colhead{${\rm Ar}$} & \colhead{${\rm Kr}$} & \colhead{${\rm Xe}$}}
\startdata
Stage 1 & $F$ & & 144.0 & 100.1 & 59.4 & 42.5 \\
& $A$      & & -8.24 & -7.27 & -5.58 & -4.55 \\
\\
Stage 2 & $F$ & & 0 & 0 & 0 & 0 \\
& $A$ & & $<-9.9$ & -6.54 & -4.67 & -5.19 \\
\\
Model & $F$ & & 144.0 & 19.0 & 6.5 & 34.6 \\
& $A$ & & -8.24 & -6.47 & -4.62 & -4.46 \\
\\
Atmosphere & $F$ & & 144.9 & 25.0 & 7.6 & 36.2 \\ 
& $A$      & & -8.21 & -6.35 & -4.52 & -4.68 \\
\enddata
\tablecomments{See Section 3 for formal definitions of $F$ and $A$. Abundances and isotopic ratios are normalized to the solar composition (see Table 1 caption for details and references). Stage 1 corresponds to a fractionation episode (generalized power law). Stage 2 corresponds to the accretion by the Earth of cometary material. The model is the superposition of these two episodes (Sections 2 and 4). The parameters that minimize the distance between the modeled and the observed composition are $(x,g,n,y,{\rm T_{\circ}})=(8.158\times 10^{-15},4.701\times 10^2,0.2614,4.407\times 10^{-7},52.36)$, see Equation 4. The results are illustrated in Figures 2 and 3.
}
\end{deluxetable}

%\clearpage

\clearpage

\begin{deluxetable}{lrccccc}
\tabletypesize{\scriptsize}
\tablewidth{0pt}
\tablecaption{Hydrodynamic Model I ($p=1$)}
\tablehead{
\colhead{} & \colhead{} & \colhead{} & \colhead{${\rm Ne}$} & \colhead{${\rm Ar}$} & \colhead{${\rm Kr}$} & \colhead{${\rm Xe}$}}
\startdata
Stage 1 & $F$ & & 161.4 & 109.1 & 67.2 & 44.4 \\
& $A$      & & -7.76 & -7.24 & -5.64 & -4.82 \\
\\
Stage 2 & $F$ & & 0 & 0 & 0 & 0 \\
& $A$ & & $<-9.9$ & -6.62 & -4.70 & -5.24 \\
\\
Model & $F$ & & 161.4 & 25.6 & 6.9 & 32.2 \\
& $A$ & & -7.76 & -6.53 & -4.65 & -4.68 \\
\\
Atmosphere & $F$ & & 144.9 & 25.0 & 7.6 & 36.2 \\ 
& $A$      & & -8.21 & -6.35 & -4.52 & -4.68 \\
\enddata
\tablecomments{See Section 3 for formal definitions of $F$ and $A$. Abundances and isotopic ratios are normalized to the solar composition (see Table 1 caption for details and references). Stage 1 corresponds to blowoff of a transient atmosphere at the end of the terrestrial accretion. Stage 2 corresponds to accretion by the Earth of cometary material. The model is the superposition of these two episodes (Section 4). Model parameters are $(s,\alpha,m^0_{cXe},p,y,{\rm T_{\circ}})=(0.9561\times 10^{-3},16.04,345,1,4.409\times 10^{-7},53.26)$, see Equations 3, 12, 13, and 14.
}
\end{deluxetable}

\clearpage

\begin{deluxetable}{lrccccc}
\tabletypesize{\scriptsize}
\tablewidth{0pt}
\tablecaption{Hydrodynamic Model II ($p\neq 1$)}
\tablehead{
\colhead{} & \colhead{} & \colhead{} & \colhead{${\rm Ne}$} & \colhead{${\rm Ar}$} & \colhead{${\rm Kr}$} & \colhead{${\rm Xe}$}}
\startdata
Stage 1 & $F$ & & 153.8 & 110.7 & 80.7 & 47.6 \\
& $A$      & & -7.98 & -7.48 & -5.68 & -4.72 \\
\\
Stage 2 & $F$ & & 0 & 0 & 0 & 0 \\
& $A$ & & $<-9.9$ & -6.64 & -4.63 & -5.20 \\
\\
Model & $F$ & & 153.8 & 17.6 & 6.5 & 35.8 \\
& $A$ & & -7.98 & -6.59 & -4.59 & -4.59 \\
\\
Atmosphere & $F$ & & 144.9 & 25.0 & 7.6 & 36.2 \\ 
& $A$      & & -8.21 & -6.35 & -4.52 & -4.68 \\
\enddata
\tablecomments{See Section 3 for formal definitions of $F$ and $A$. Abundances and isotopic ratios are normalized to the solar composition (see Table 1 caption for details and references). Stage 1 corresponds to blowoff of a transient atmosphere at the end of the terrestrial accretion. Stage 2 corresponds to accretion by the Earth of cometary material. The model is the superposition of these two episodes (Section 4). Model parameters are $(s,\alpha,m^0_{cXe},p,y,{\rm T_{\circ}})=(2.841\times 10^{-5},12.89,140,3.980,5.400\times 10^{-7},54.55)$, see Equations 3, 12, and 14.
}
\end{deluxetable}

\begin{deluxetable}{lccc}
\tabletypesize{\scriptsize}
\tablewidth{0pt}
\tablecaption{Major Volatiles}
\tablehead{
\colhead{} & \colhead{H} & \colhead{C} & \colhead{N}}
\startdata
Cometary Accretion & $1.6\times 10^{20}-2.5\times 10^{20}$ & $3.4\times 10^{19}-5.1\times 10^{19}$ & $4.8\times 10^{18}-1.3\times 10^{19}$ \\
Terrestrial Inventory & $1.9\times 10^{23}-1.3\times 10^{24}$ & $2.1\times 10^{23}-2.1\times 10^{24}$ & $4.3\times 10^{20}-7.8\times 10^{20}$ \\
\enddata
\tablecomments{Amounts are expressed in mol. The mass of comets accreted by the Earth is $3\times 10^{18}$ kg (${\rm M_o}=y{\rm M}_{\oplus}$). The cometary concentrations for major volatiles are taken from Delsemme (1988) and Marty and Dauphas (2002). The terrestrial inventories refer to the bulk Earth excluding the core and are taken from Dauphas and Marty (2002).
}
\end{deluxetable}

\clearpage

\begin{deluxetable}{lrccccc}
\tabletypesize{\scriptsize}
\tablewidth{0pt}
\tablecaption{Mantle Model}
\tablehead{
\colhead{} & \colhead{} & \colhead{} & \colhead{${\rm Ne}$} & \colhead{${\rm Ar}$} & \colhead{${\rm Kr}$} & \colhead{${\rm Xe}$}}
\startdata
Stage 1 & $F$ & & 0 & 0 & 0 & 0 \\
& $A$      & & -10.11 & -10.11 & -10.11 & -10.11 \\
\\
Stage 2 & $F$ & & 177.5 & 26.8 & 7.4 & 1.0 \\
& $A$ & & $-12.48$ & -10.52 & -8.92 & -7.72 \\
\\
Stage 3 & $F$ & & 144.9 & 25.0 & 7.6 & 36.2 \\
& $A$ & & $-10.50$ & -8.64 & -6.81 & -6.97 \\
\\
Model & $F$ & & 53.7 & 24.3 & 7.6 & 30.9 \\
& $A$ & & -9.96 & -8.62 & -6.80 & -6.90 \\
\\
Mantle & $F$ & & 47.1 & 25.0 & 7.6 & 31.8 \\ 
& $A$      & & -10.33 & -8.76 & -6.77 & -6.41 \\
\enddata
\tablecomments{See Section 3 for formal definitions of $F$ and $A$. Abundances and isotopic ratios are normalized to the solar composition (see Table 1 caption for details and references). Stages 1 and 2 represent the entrapment of solar and meteoritic gases in the mantle. Stage 3 corresponds to recycling of atmospheric volatiles in the silicate Earth. The model is the superposition of these three episodes. Model parameters are $(x,y,z)=(7.730\times 10^{-11},1.853\times 10^{-6},5.158\times 10^{-3})$, see Section 6, Equation 15. Note that during recycling, atmospheric volatiles would have been elementally fractionated, and stage 3 may be modified accordingly. The results are illustrated in Figures 4 and 5.
}
\end{deluxetable}

\newpage

\clearpage
\begin{figure}
\epsscale{0.5}
\plotone{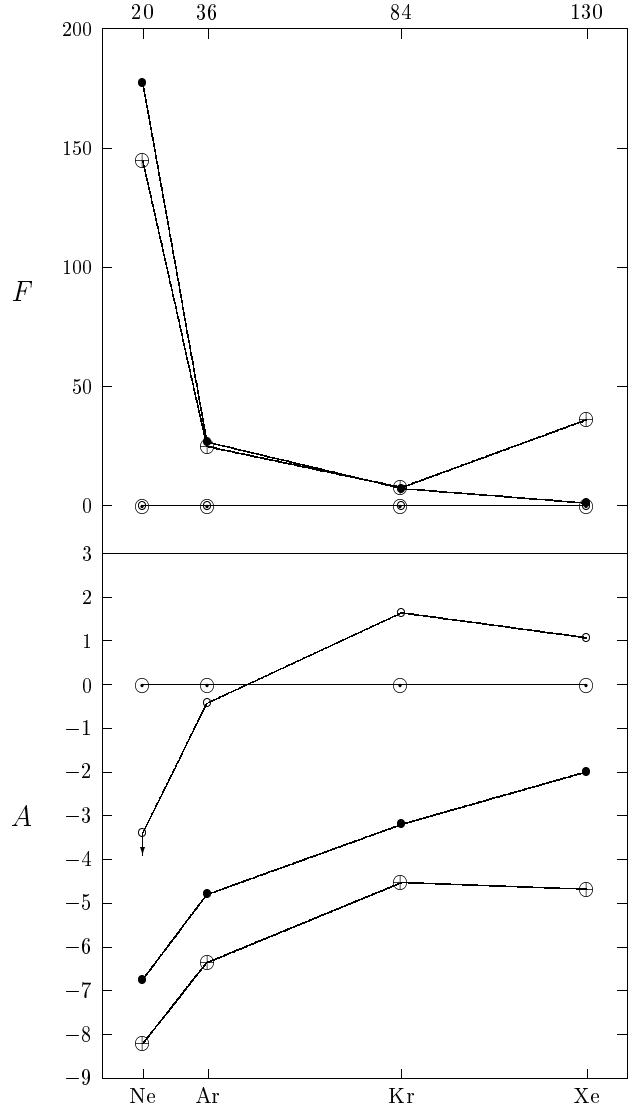}
\caption{\scriptsize Normalized isotopic ($F$) and elemental ($A$) composition of the sun ($\odot$), the terrestrial atmosphere ($\oplus$), asteroids ($\bullet$), and comets formed at 55 K ($\circ$). See Section 3 and Table 1 for details and references. One would expect heavy noble gases to be less depleted and fractionated than light ones. This is not true for xenon in the terrestrial atmosphere. Despite being heavier, xenon is more depleted and more fractionated than krypton. This is known as the "missing xenon" paradox.}
\end{figure}

\newpage

\clearpage
\begin{figure}
\epsscale{0.5}
\plotone{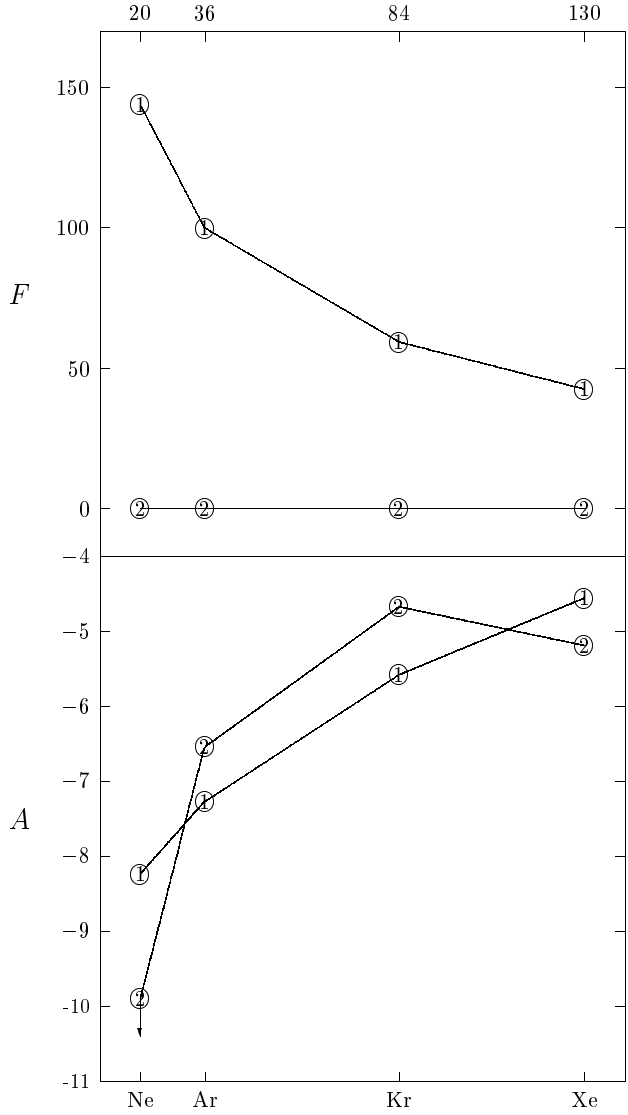}
\caption{\scriptsize The normalized isotopic ($F$) and elemental ($A$) compositions are reported for the two stages of atmospheric formation invoked in the Phenomenological Model (Sections 2 and 4, and Table 2). The fractionation episode was modeled using the generalized power law. The parameter set that minimizes the distance between the modeled and the observed atmosphere is  $(x,g,n,y,{\rm T_{\circ}})=(8.158\times 10^{-15},4.701\times 10^{2},0.2614,4.407\times 10^{-7},52.36)$. It is assumed that at the end of Earth's accretion, continuous loss of a transient gas envelope resulted in elemental and isotopic fractionation of the residual atmosphere (stage 1). Later or at the same time, the Earth accreted cometary material having fractionated elemental abundances but unfractionated isotopic ratios (stage 2). The resulting atmosphere would have near solar Xe/Kr ratio, almost unfractionated krypton delivered by cometary material, and fractionated xenon inherited from the fractionation episode. Such a scenario therefore provides a satisfactory solution to the "missing xenon" paradox.}
\end{figure}

\newpage

\clearpage
\begin{figure}
\epsscale{0.5}
\plotone{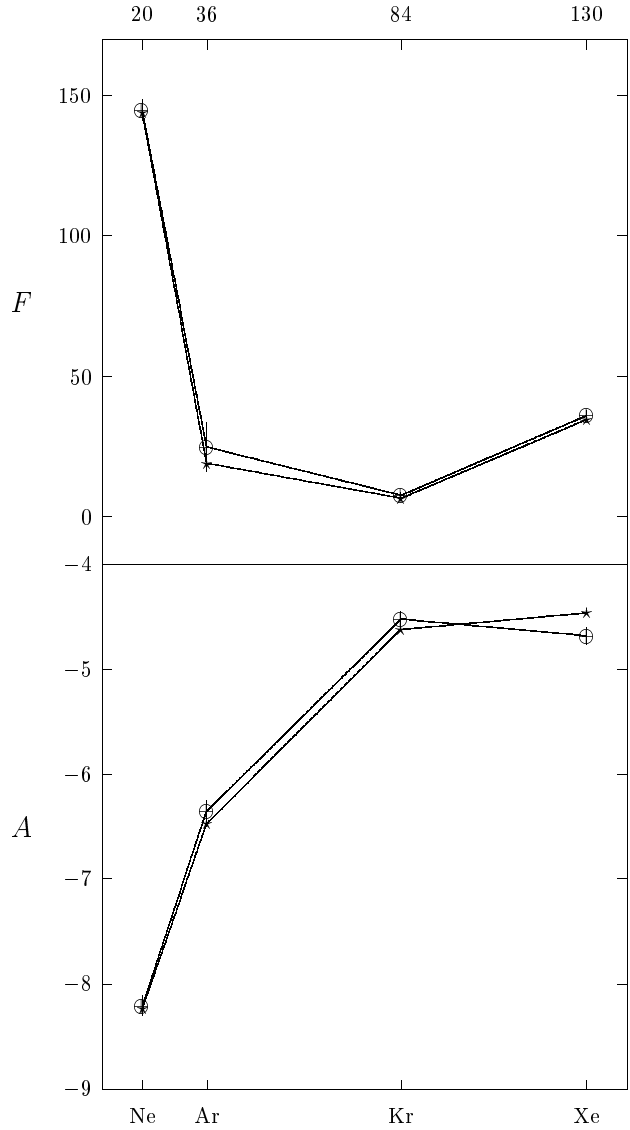}
\caption{\scriptsize Isotopic ($F$) and elemental ($A$) compositions of the observed ($\oplus$) and the simulated ($\star$) atmosphere in the Phenomenological Model. The modeled atmosphere is the superposition of the two stages illustrated in Fig. 2 (Sections 2 and 4, Equation 4, and Table 2). The difference between the observed and the simulated atmosphere can be ascribed to inadequacy of the fractionation law and uncertainties regarding the compositions of comets and the nebula. Uncertainties are $2\sigma$.}
\end{figure}

\newpage

\clearpage
\begin{figure}
\epsscale{0.5}
\plotone{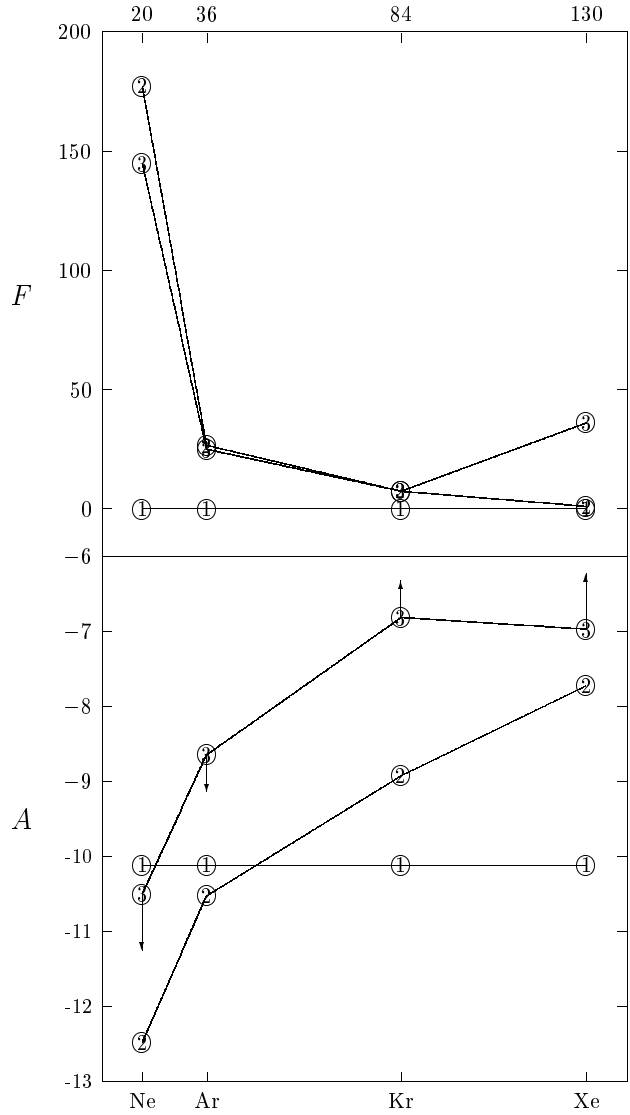}
\caption{\scriptsize  The normalized isotopic ($F$) and elemental ($A$) compositions are reported for the three stages invoked in the Mantle Model (Section 6 and Table 6).  Stages 1 and 2 represent the entrapment of solar and possibly meteoritic gases in the mantle. Stage 3 corresponds to recycling of atmospheric volatiles in the silicate Earth. The set of parameters that minimizes the relative distance between the simulated and the observed pattern is $(x,y,z)=(7.730\times 10^{-11},1.853\times 10^{-6},5.158\times 10^{-3})$. The arrows indicate the expected effect of fractionation during recycling.}
\end{figure}

\newpage

\clearpage
\begin{figure}
\epsscale{0.5}
\plotone{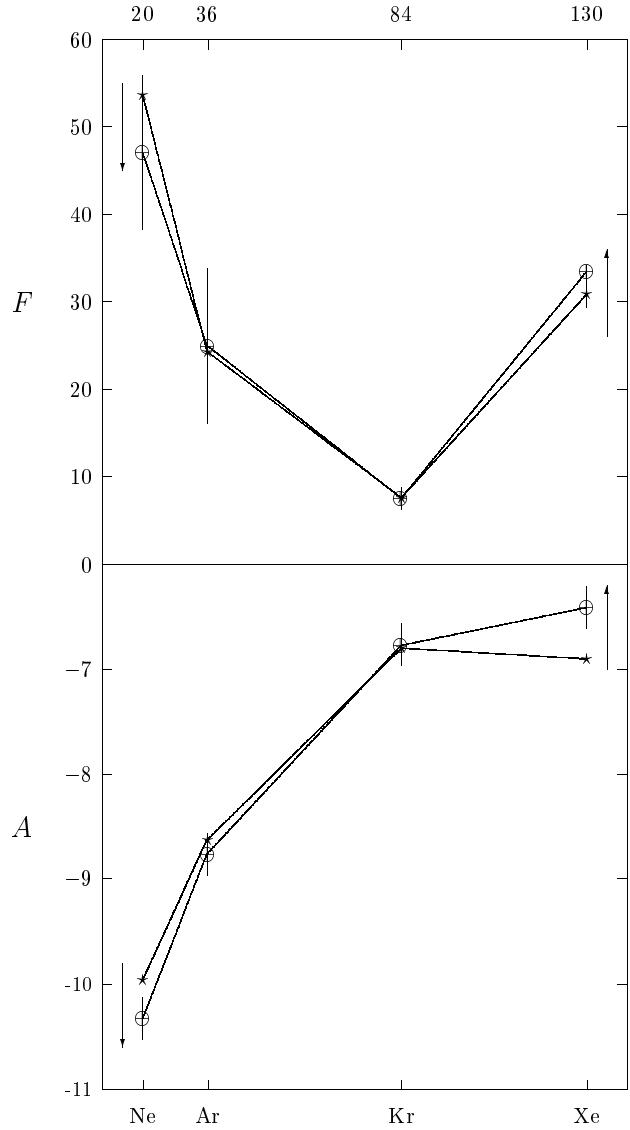}
\caption{\scriptsize  Isotopic ($F$) and elemental ($A$) compositions of the observed ($\oplus$) and the simulated ($\star$) rare gas pattern in the Mantle Model. The modeled mantle is the superposition of the three stages illustrated in Fig. 4 (Section 6 and Table 6). The arrows indicate the direction the modeled composition would be affected if atmospheric noble gases had been fractionated during recycling, heavy elements being less depleted than light ones. Uncertainties are 2$\sigma$.}
\end{figure}

\newpage

\clearpage
\begin{figure}
\epsscale{0.9}
\plotone{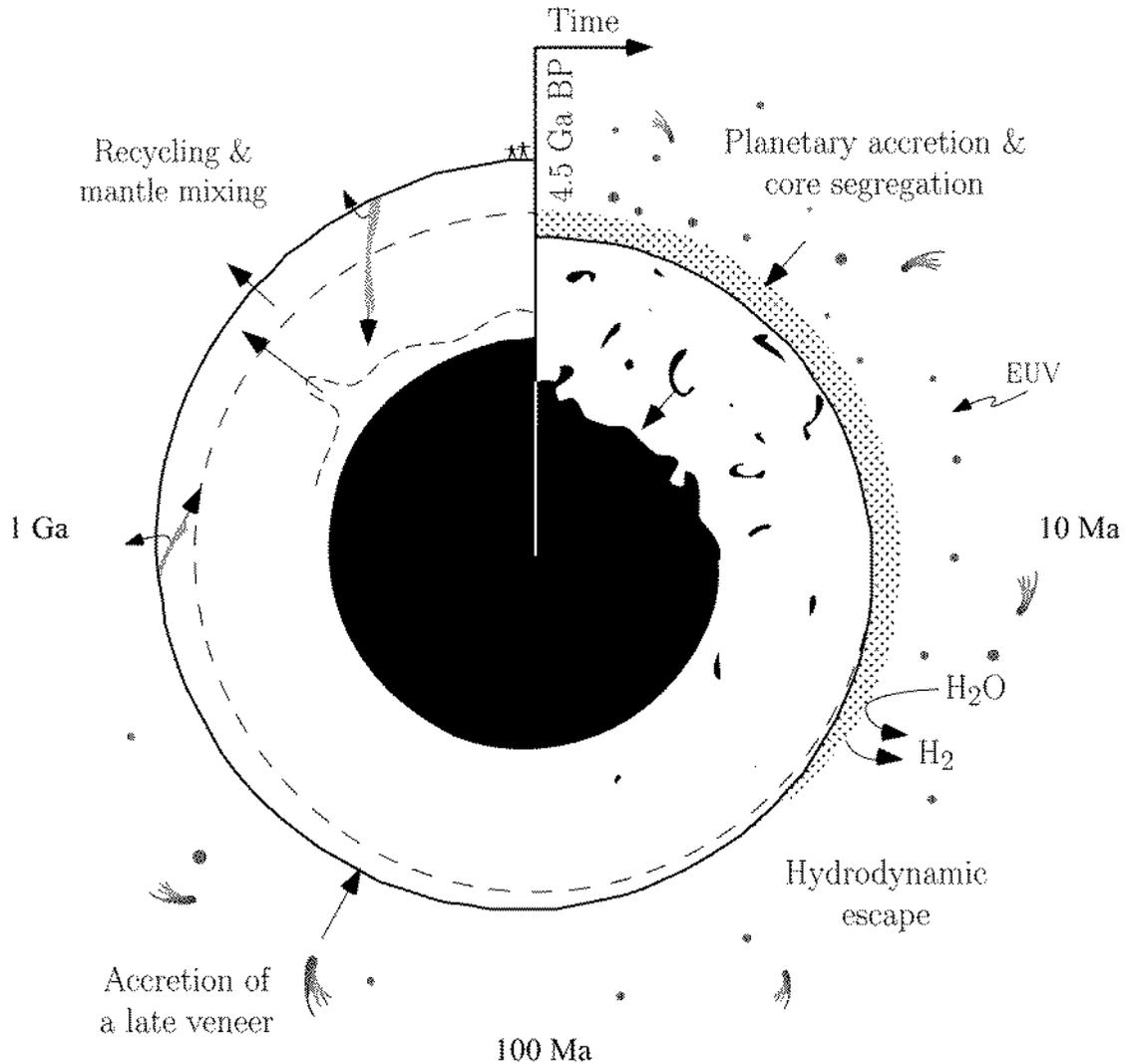}
\caption{\scriptsize Cartoon illustrating the preferred scenario for the formation of the terrestial atmosphere. The silicate Earth incorporated solar and possibly meteoritic gases during its formation. Escape to space of hydrogen trapped directly from the nebula or derived from the reduction or photodissociation of water permitted the loss of heavy noble gases through hydrodynamic escape. Major volatiles which form reactive compounds with the silicate Earth were more efficiently retained by the growing planet. Within 30 Ma of the solar system formation, the core completed its formation. Extraterrestrial material accreted after that time did not equilibrate with the core and is referred to as the late veneer. Noble gases delivered by the comets accreted during this period were mixed with volatiles remaining after the escape episode. During geological times, elementally fractionated atmospheric volatiles were recycled in the mantle. The recycled atmospheric gases were mixed with gases trapped in the deep Earth during its accretion. All these successive stages may have overlapped. Timescales are indicated.}
\end{figure}

\end{document}